\documentclass[journal = jctcce,layout=twocolumn]{achemso}

\usepackage{color}
\usepackage[numbers,super,sort&compress]{natbib}
\usepackage{amssymb}
\usepackage{booktabs} 
\usepackage[T1]{fontenc}
\usepackage{achemso}
\usepackage[ngerman,english]{babel}
\usepackage{amsfonts}
\usepackage{amsmath}
\usepackage{array}
\usepackage{float}
\usepackage[version=3]{mhchem}
\usepackage{textcomp}
\usepackage{placeins}
\usepackage{footnote}
\usepackage[para]{threeparttablex}
\makesavenoteenv{tabular}
\makesavenoteenv{table}
\newcolumntype{x}[1]{>{\centering\arraybackslash\hspace{0pt}}p{#1}}

\author{Olga S. Bokareva}
\author{Gilbert Grell}
\author{Sergey I. Bokarev}
\email{sergey.bokarev@uni-rostock.de}
\author{Oliver K\"{u}hn}

\affiliation{
	Institut f\"{u}r Physik, Universit\"{a}t Rostock, 
	Universit\"{a}tsplatz 3, D-18055 Rostock, Germany
}
\title{Tuning Range-Separated Density Functional Theory for Photocatalytic Water Splitting Systems}

\begin{document}
%Macros are defined here
\newcommand{\Fig}[1]{Fig.~\ref{#1}}
\newcommand{\eq}[1]{eq.~\ref{#1}}
\newcommand{\com}[1]{\textcolor{red}{#1}}
\newcommand{\hl}[1]{\textcolor{blue}{#1}}

\begin{abstract}
We discuss the system-specific optimization of long-range separated density functional theory (DFT) for the prediction of electronic properties relevant for a photocatalytic cycle based on an Ir(III) photosensitizer (IrPS). Special attention is paid to the charge-transfer properties, which are of key importance for the photoexcitation dynamics, but   and cannot be correctly described by means of conventional DFT. The optimization of the range-separation parameter using the $\Delta$SCF method is discussed for IrPS including its derivatives and complexes with electron donors and acceptors used in photocatalytic hydrogen production. Particular attention is paid to the problems arising for a description of medium effects by means of a polarizable continuum model.
\end{abstract}

\maketitle
%
%-------------------------------------------------------------
\section{Introduction}

%-------------------------------------------------------------
Transition metal (TM) organometallic complexes have found a wide  range of catalytic, medical, and biological applications.  In view of the growing demand in energy, one of the  most perspective applications of TM complexes is their role in systems for conversion and storage of solar light energy into chemical form.  Here, a variety of  homogeneous and heterogeneous schemes have been suggested.~\cite{Esswein-cr-2007, Hambourger-csr-2009, Blankenship-s-2011} TM complexes are attractive because of their notable spin-orbit coupling.  This facilitates the absorption of  sun light, whose energy can be stored in charge-separated triplet states. These long-living triplet states are then available for further reactions. Besides light-harvesting properties, TM  complexes are also used as direct catalysts for water splitting; for  recent reviews on homogeneous photocatalysis see, e.g. refs\, \cite{Hagfeldt-cr-2010, Eckenhoff-dt-2012}.

To develop new efficient and stable photocatalytic  systems,  understanding of their primary photoreaction steps and especially excited state properties is required.~\cite{You-csr-2012} At the moment, most computationally demanding studies of sizeable TM complexes applied in catalysis are performed  with DFT in combination with the B3LYP functional and its extension in the time-domain in the linear response formulation (TDDFT).~\cite{Tinker-c-2007, Li-pccp-2009, Costa-jms-2009, Tian-ejic-2010, Ladouceur-ic-2010, Wu-jpca-2010, Takizawa-ic-2014, Chirdon-ic-2014, Song-sa-2014} Besides efficiency it is the absence of system-dependent parameters that need to be determined first (such as the active space in multi-reference methods), which makes (TD)DFT attractive. The power of the DFT method to reproduce different electronic ground state properties  even for rather large systems is well documented.~\cite{Sousa-jcpa-2007} Concerning excited state properties of TM complexes, however, there appears to be no unequivocal opinion; for a  review see ref\,\cite{Laurent-ijqc-2013}. For an Iridium (III) heteroleptic complex, for instance, we have shown that the excited state energies are strongly dependent on the employed exchange-correlation functional.~\cite{Bokarev-jcp-2012}  Nevertheless,  the TDDFT approach remains to be the most attractive one.

Many of the practical problems are related to the erroneous description of charge transfer (CT) states.~\cite{Dreuw-cr-2005, Peach-jcp-2008} To overcome this drawback, different schemes  have been proposed, such as, e.g., scaled hybrids  \cite{Becke-jcp-1993a, Adamo-jcp-1998} and range-separated hybrid functionals.~\cite{Savin1995, Leininger-cpl-1997}

Benchmarking DFT results for  TM complexes in solution  is hampered by the facts that  there are no high-resolution experimental data available and only a limited number of theoretical approaches might be applicable to generate a presumably more accurate reference. At the moment, multi-reference perturbation theory is  the best choice for reference calculations of excited state properties.  However, in multi-reference approaches the problem of a sufficiently large and balanced active space has to be addressed  for every system.~\cite{Bokarev-jcp-2012}

\begin{figure*}[t]
\includegraphics[width=0.95\textwidth]{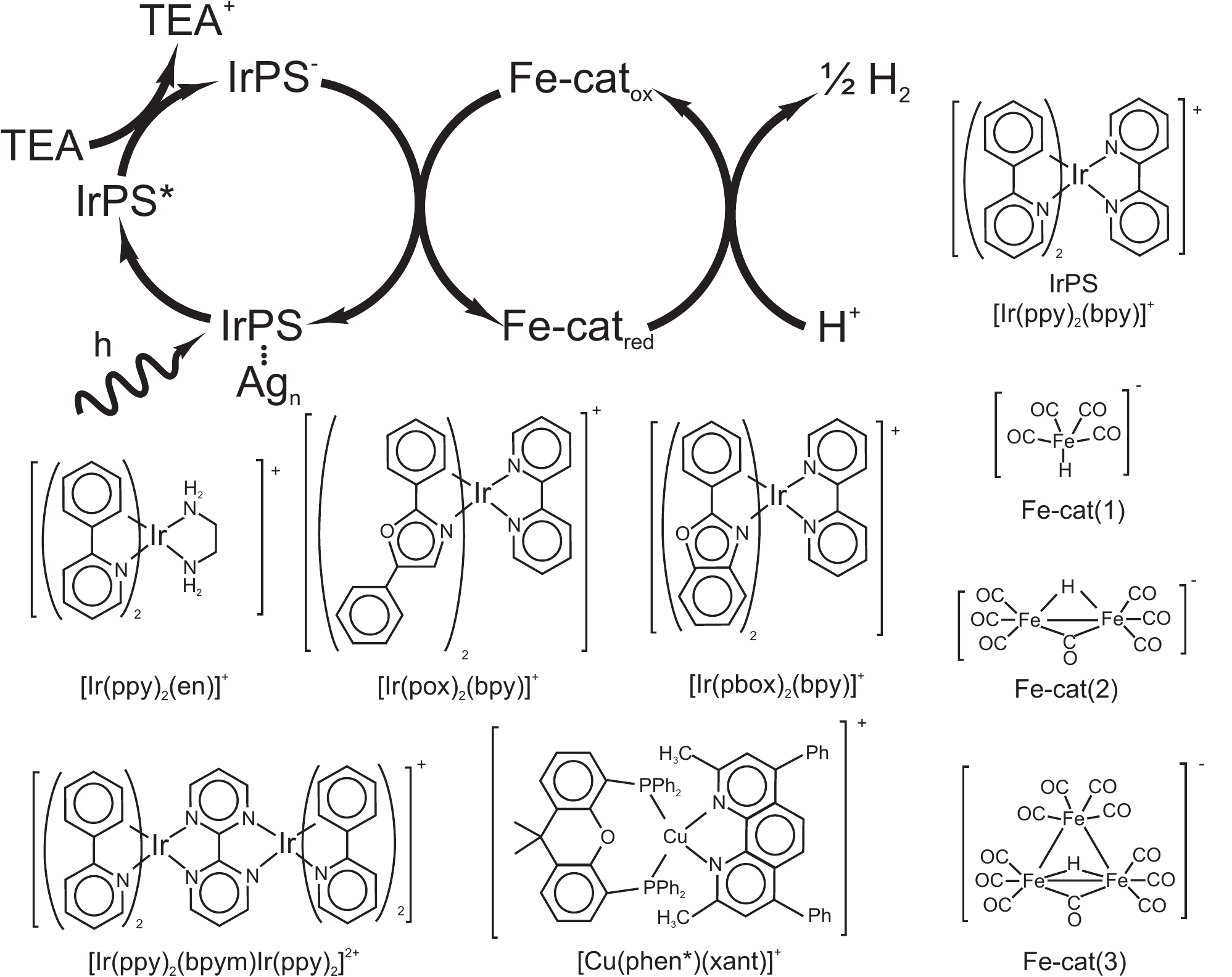}
\caption{\label{First_scheme}
Scheme of photocatalytical water splitting introduced in ref\,~\cite{Gaertner-acie-2009} and  structural formulas of photosensitizers and water  reduction catalysts studied in this work.}	
\end{figure*}

In this paper, we focus on the tuning of the range  separation parameter for the long-range corrected LC-BLYP functional as applied for a particular photocatalytic system developed  by Beller et al.~\cite{Gaertner-acie-2009} It  consists of a heteroleptic Ir(III) photosensitizer, triethylamine  (TEA) as a sacrificial reductant and a series of iron  carbonyls (Fe-cat) as water reduction catalyst (Figure 1). 
%\ref{First_scheme}
In addition, hybrid systems of IrPS and  silver nanoparticles are considered.~\cite{Bokareva-pccp-2012, Bokareva-cp-2014, Bokareva-pccp-2014-prep} Finally, the scope is broadened by including results on modified IrPSs and copper based PS which could replace noble-metal containing PS, see ref. \cite{Fischer-c-2014} and references therein. Although the choice of the system is arbitrary, its general properties as well as the basic reaction steps are common for TM photocatalysis. Therefore, the present study provides some general insight into the role of tuning the range separation parameter in these systems.

In the following section, we start  with a brief introduction in the theory of long-range separated functionals.  We then present our results on  tuning the range-separation functional for IrPS and joint IrPS-X  systems, where X is a reactant responsible for a particular process, i.e. light absorption, electron transfer and recovery  of the ground state. The influence of the reference  geometry, environmental and basis set effects are discussed.  In particular, the applicability of the $\omega$-tuning together with the Polarizable Continuum Model (PCM) is critically analyzed.

%-------------------------------------------------------------
\section{Optimally Tuned Range-Separated Functionals}
\label{sec:Tuning}
%-------------------------------------------------------------

DFT has its formal roots in the theorems of Hohenberg and Kohn \cite{Hohenberg-pr-1964, Kohn-pr-1965} and it can be considered as an in principle exact approach. In practice, it is hampered by the lack of the exact exchange correlation (XC) functional, $E_{\rm XC}[\rho(\vec{r})]$, with   $\rho(\vec{r})$ being the electron density. There seems to be no first principle route to $E_{\rm XC}[\rho(\vec{r})]$ and therefore it is usually constructed on the basis of model systems or fitted   to experimental reference data. Excited state calculations performed by TDDFT in the linear response approximation employ XC functionals  obtained for the electronic ground state.
 
 All caveats and the sometimes spurious behaviour of the DFT approach originate from the approximation of the unknown XC kernel.  There are  three main problems: First, in the  $r \rightarrow \infty$ limit the potential and hence the density itself possess an incorrect behaviour.  In principle,  at large distances  the potential should be dominated by the exchange  term decaying as $-1/r$,  while the correlation term decays as $\approx -1/r^{4}$.~\cite{Handy-pr-1969, Almbladh-prb-1985}  However, for the local density approximation and to a lesser extent for the generalized gradient approximation (GGA), the exchange term decays  exponentially.~\cite{Baer-arpc-2010} Second, the approximate nature of the exchange term does not cancel the Coulomb interaction of the electron with itself at large $r$ as it is the case for the Hartree-Fock (HF) method (self-interaction error). Third, a well-known deficiency of approximate Kohn-Sham approach is  that the fundamental gap, defined as a difference between ionization potential (IP) and electron affinity (EA), differs notably from the orbital energy difference $\epsilon_{\rm{HOMO}} - \epsilon_{\rm{LUMO}}$.~ \cite{Kummel-rmp-2008}  This can be explained by the finite jump of the Kohn-Sham correlation  potential for a statistic ensemble with variable number of electrons while passing through integer ($N$) number of electrons,  the so-called derivative discontinuity.~  \cite{Perdew-prl-1982, Sham-prl-1983, Godby-prl-1986, Allen-mp-2002, Kummel-rmp-2008} According to Koopmans' theorem  for the DFT case,~\cite{Almbladh-prb-1985, Perdew-prl-1982, Cohen-prb-2008, Salzner-jcp-2009} the HOMO corresponds to the IP, but the LUMO is generally more strongly bound than in HF theory and cannot be related to the EA.

Considering TDDFT calculations of electronic excitation energies, the error  correlates with the overlap of the donor and acceptor orbitals, being the largest for CT  and Rydberg states.~ \cite{Dreuw-cr-2005, Peach-jcp-2008} For vanishing overlap between these orbitals (e.g., due to long-range CT), the excitation energy is reduced to the orbital energy difference, being a poor estimate in case of DFT, which is in contrast to HF theory.

The remedy for the erroneous exchange potential could be  the substitution of the approximate density-dependent  exchange energy with the exact orbital-dependent one \cite{Seidl-prb-1996} within generalized Kohn-Sham theory.~\cite{Seidl-prb-1996, Kummel-rmp-2008, Baer-arpc-2010} The exact exchange can be included in a fixed manner  like in hybrid functionals (e.g. B3LYP) or weighted  with a function depending on the inter-electron distance $r_{12}$. The latter case is implemented in range-separated  functionals via splitting the Coulomb operator into  local and non-local parts
\begin{equation}\label{LCeq} 
\frac{1}{r_{12}}=\frac{1-\Gamma(\omega r_{12})}{r_{12}}+\frac{\Gamma(\omega r_{12})}{r_{12}},
\end{equation}
where $\Gamma(\omega r_{12})$ is a smooth range-separation function, which damps the exchange contribution  from the density functional and complements it  with exact exchange. Examples are the Yukawa kernel \cite{Baer-prl-2005, Akinaga-cpl-2008} $e^{-\omega r}/r$  or the error function \cite{Iikura-jcp-2001, Leininger-cpl-1997} kernel ${\rm erf}(\omega r)/r$. Such an approach eliminates the spurious behaviour  because the exact exchange has the correct asymptotic  character and cancels the self-interaction exactly.~\cite{Gerber-cpl-2005, Zhao-jpca-2006,  Peach-pccp-2006, Livshits-pccp-2007, Chai-jpc-2008, Mori-jcp-2006} 

Certain standard values for the range-separation parameter $\omega$ have been established in refs\,~\cite{Tawada-jcp-2004, Song-jcp-2007} and they are used like universal constants in popular quantum chemical programs.~\cite{Schmidt-jcc-1993,g09} The value $0.33 \, \text{bohr}^{-1}_{}$ was  determined by a least square  fit to empirical data for first- to  third-row atoms \cite{Tawada-jcp-2004} and  later refined to $0.47 \, \text{bohr}^{-1}_{}$  for larger sets of small molecules.~\cite{Song-jcp-2007}  (for other test sets, see also refs.\,~\cite{Yanai-cpl-2004,  Livshits-pccp-2007, Chai-jpc-2008, Rohrdanz-jcp-2009}) However, it is clear that for molecules, which have not been part of the training sets, a more accurate description will be provided by choosing a system-dependent $\omega$.  An optimal $\omega$ for a particular system can be determined by ensuring that the energy of  the HOMO orbital equals the negative of the IP, a relation that would be fulfilled for the exact functional.~\cite{Perdew-prb-1997}

However, in the following we will use alternative the so-called  $\Delta {\rm SCF}$ method,~\cite{Livshits-pccp-2007, Stein-jacs-2009, Stein-jcp-2009}  where the IP and EA are calculated as the differences  between ground state (gs) energies of systems with $N$ and $N\pm1$ electrons, i.e.

\begin{equation}\label{eq:IP}
{\rm IP}^{\omega}_{}(N)=E_{\rm gs}^{\omega}(N-1)-E^{\omega}_{\rm gs}(N) \, ,
\end{equation}

\begin{equation}
{\rm IP}^{\omega}_{}(N+1)={\rm EA}^{\omega}_{}(N)=E_{\rm gs}^{\omega}(N)-E^{\omega}_{\rm gs}(N+1) \, .
\end{equation}

This yields the separate tuning conditions

\begin{equation} \label{J} 
J^{}_{0}(\omega)=\left|\varepsilon^{\omega }_{\rm HOMO}(N)+\text{IP}^{\omega}_{}(N)\right| \, ,
\end{equation}

\begin{equation} \label{J1} 
J^{}_{1}(\omega)=\left|\varepsilon^{\omega }_{\rm HOMO}(N+1)+\text{EA}^{\omega}_{}(N)\right| \, . 
\end{equation}

In order to obtain a proper description of the  fundamental gap, the functions $J^{}_{0}(\omega )$ and $J^{}_{1}(\omega )$  for IP and EA should be minimized simultaneously. For the present one-parameter formulation this requires to minimize the general function 
\begin{equation}\label{Jgeneral}
J^{}_{}(\omega) = J^{}_{0}(\omega)+J^{}_{1}(\omega) \, .
\end{equation}
Note that in general $J(\omega)$ can be non-zero even for exact functionals, because  it is defined for systems with different numbers of electrons.

Although minimizing $J^{}_{}(\omega )$   was applied extensively for tuning the range-separation parameter, in ref\,~\cite{Karolewski-jcp-2013} it was stressed that the $\omega$-values are very  close to minima of $J_{0}^{}(\omega)$, eq 4, 
%(\eq{J}) 
or 
$J_{1}^{}(\omega)$, eq 5, 
%(\eq{J1}) 
depending on the particular form of the dependencies of  $\varepsilon_{\rm HOMO}^{}$ and the  total energy on $\omega$. To avoid such a biased behaviour, it was suggested to apply a least-square approach to minimize  the resulting error according to \cite{Kronik-jctc-2012,  Karolewski-jcp-2013} 
\begin{equation}\label{Jstar}
J^{*}_{}(\omega)=\sqrt{J^{2}_{0}(\omega)+J^{2}_{1}(\omega)}  \, .
\end{equation}

Alternatively, for long-range CT systems where donor and acceptor units can be clearly distinguished, ground states of neutral donor and negatively ionized acceptor can be considered  in a tuning approach \cite{Stein-jcp-2009, Stein-jacs-2009} based on Mulliken's rule.~\cite{Mulliken-jacs-1950}

\begin{figure*}[t]
	\includegraphics[width=1.0\textwidth]{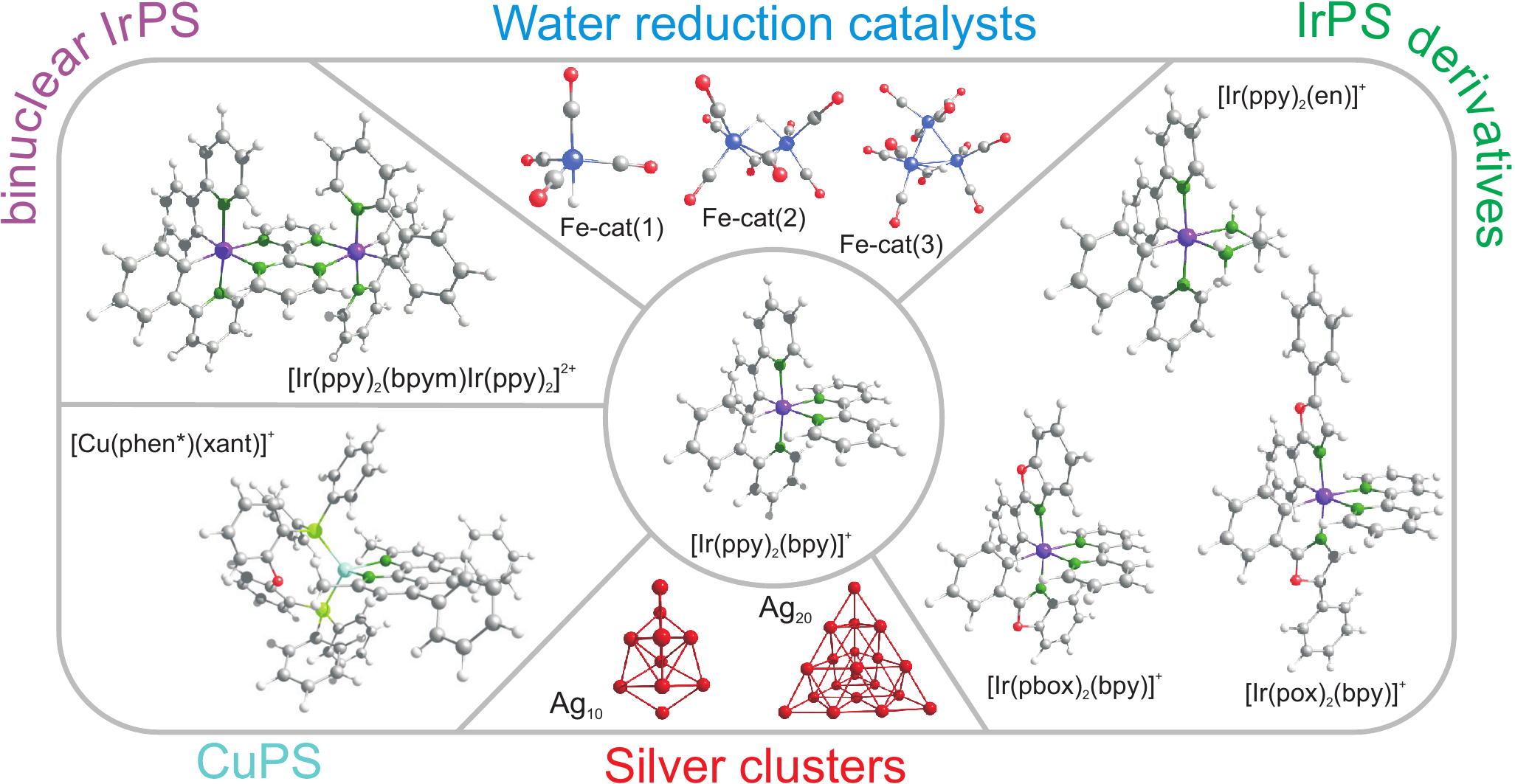}
	\caption{\label{Structures}
		Overview of all optimized structures of those molecules that were used for $\omega$-tuning.}	
\end{figure*}

It is important to stress that the optimal $\omega$  is density and thus system dependent.  Thereby, the dependence of $\omega$ on the density for range-separated  functionals is much more pronounced than that of the scaling factor for hybrid functionals.~\cite{Baer-arpc-2010} Since the range-separated functionals eliminate self-interaction \cite{Gerber-cpl-2005, Zhao-jpca-2006, Vydrov-jcp-2006,  Peach-pccp-2006, Livshits-pccp-2007, Chai-jpc-2008, Mori-jcp-2006} and have the correct asymptotic form of  non-local exact exchange,  the correct Coulomb-like  behaviour of the energy is assured and thus  CT and Rydberg excitation energies are improved. Additionally, since the tuned range-separation parameter  improves the fundamental gap for the $N$ electron system, one can expect also a better description of electronic excitation energies in TDDFT, where the leading term reads as $\epsilon_{\rm HOMO}-\epsilon_{\rm LUMO}$.

In the past it was shown in numerous publications the the optimally  tuned functionals indeed improve the description  of different molecular properties such as IPs, fundamental and optical gaps,  and CT and Rydberg transition energies.~\cite{Seidl-prb-1996, Tawada-jcp-2004, Sekino-jcp-2007, Cohen-prb-2008,  Lange-jpcb-2008, Lange-jpcb-2008a, Chai-jpc-2008, Peach-jcp-2008, Stein-jcp-2009, Stein-jacs-2009, Akinaga-ijqc-2009, Stein-prl-2010,  Refaely-prb-2011, Kummel-rmp-2008, Wong-pccp-2009, Sears-jcp-2011, Korzdorfer-jcp-2011, Karolewski-jcp-2013,  Kuritz-jctc-2011, Moore-co-2012, Egger-jctc-2014, Karolewski-jcp-2011, Song-jcp-2007, Foster-jctc-2012, Srebro-jctc-2012, Srebro-jpcl-2012, Minami-ijqc-2013, Autschbach-acr-2014} The accuracy of optimally tuned range-separated  functionals was critically tested in ref\,~\cite{Karolewski-jcp-2013} for basis set variation as well as prediction  of relative energies of spin states, binding energies,  and the form of potential energy surfaces. 

Finally, we note that a generalized form of LC-partitioning of Coulomb interaction proposed by Yanai for CAM-B3LYP \cite{Yanai-cpl-2004} is often applied for non-empirical optimal tuning as well (see, e.g. refs\,~\cite{Srebro-jctc-2012, Refaely-prl-2012, Egger-jctc-2014, Srebro-jpcl-2012, Minami-ijqc-2013, Autschbach-acr-2014}). Additionally, the  simultaneous two-parameter optimization of short- and long-range separation parameters was proposed in order to minimize the delocalization error. Successfully applications provided an improved description of molecular properties.~\cite{Srebro-jctc-2012, Srebro-jpcl-2012, Autschbach-acr-2014}

%-------------------------------------------------------------
\section{Computational Details}
\label{sec:Comp}
%-------------------------------------------------------------
%
The tuning of the range-separation parameter $\omega$ according to Eqs 6 and 7
%\ref{Jgeneral} and \ref{Jstar} 
has been  done for the LC-BLYP functional.~\cite{Iikura-jcp-2001,Tawada-jcp-2004, Chiba-jcp-2006} It  includes the Becke exchange  \cite{Becke-pra-1988} and a correlation part of GGA-type established by Lee, Yang and Parr.~\cite{Lee-prb-1988, Miehlich-cpl-1989} This scheme uses 
%Eq. \ref{LCeq} 
eq 1 with an error function  kernel as suggested by Hirao et. al..~\cite{Iikura-jcp-2001} For the  case of the parent IrPS compound, the results with a tuned $\omega$ are  compared to those obtained with the standard  system-independent values of $0.33\:\text{bohr}^{-1}_{}$ and  $0.47\:\text{bohr}^{-1}_{}$ as well as  with BLYP  and B3LYP  \cite{Becke-jcp-1993} functionals.

All calculations have been performed  using the LANL2DZ ECP basis set for Ir, Fe, Cu, and Ag  and the 6-31G(d) basis set for all other atoms.  In ref\,~\cite{Karolewski-jcp-2013} it was pointed to  a strong dependence of the optimal  $\omega$ on the inclusion of diffuse function in case of  simple atomic and diatomic systems. To test the influence of diffuse basis functions for the present example of IrPS, calculations have been performed with 6-31+G(d) and 6-31++G(d) basis sets. The inclusion of one or two diffuse function has led to decrease of $\omega$ by $0.01\:\text{bohr}^{-1}_{}$. Since this effect appears to be of minor importance, below only results for the 6-31G(d) basis set will be presented.

The impact of reference geometry used for tuning of $\omega$ was studied on the example of IrPS.  The tuning for the slightly different geometries, which have been obtained  within LC-BLYP with the two standard (0.33 and 0.47) and the optimized (0.18) $\omega$, led to the  same value of $0.18\:\text{bohr}^{-1}_{}$.  In the following, the structures optimized with LC-BLYP  with optimal $\omega$ have  been utilised for tuning (see Figure 2). 
%\ref{Structures}
For the complexes \ce{IrPS-X} (X=TEA or Fe-cat)  the geometries of the constituing parts have been first  optimized separately and then placed at fixed  positions without further optimization (for further details see ref\,~\cite{Neubauer-jpcl-2014}). For systems containing silver clusters,  the geometries were optimized in  refs\,~\cite{Bokareva-pccp-2012, Bokareva-cp-2014, Bokareva-pccp-2014-prep}.  The standard TDDFT formalism was applied for excited state calculations. Solvent effects have been included within the PCM approach.~\cite{Tomasi-cr-2005} All calculations were done with the Gaussian09 set of programs.~\cite{g09}

%
%-------------------------------------------------------------
\section{Results and Discussion}
\label{sec:Res}
%-------------------------------------------------------------
% 
\subsection{Optimization of the Range-Separation Parameter}
\begin{figure*}[t]
	\includegraphics[width=1.1\textwidth]{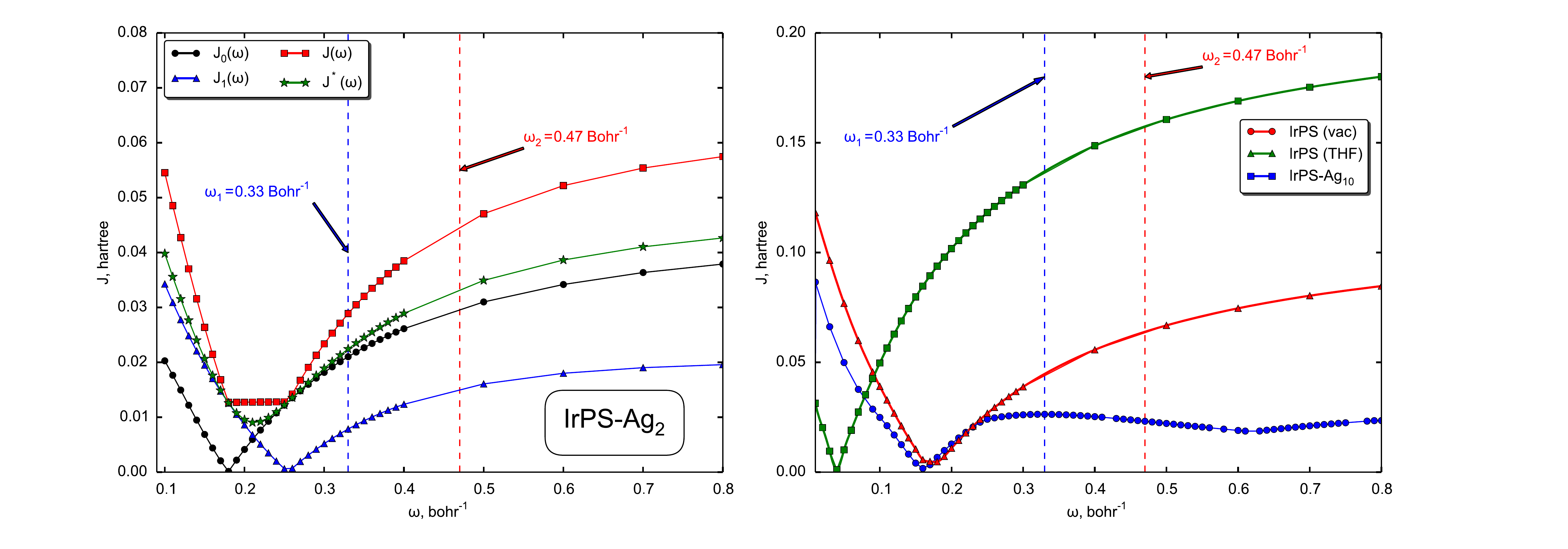}
	\caption{\label{Curves}
		Left panel: The functions defined in eqs 4--7 used for optimization of the range-separation parameter $\omega$ 		for \ce{IrPS-Ag2}. Right panel: Selected examples of 
		$J(\omega)$  for IrPS in different environments.}	
\end{figure*}

The optimal values of the range-separation parameter $\omega$ for molecules and complexes related to the studied photocatalytic system for hydrogen generation (Figure 1) 
%\ref{First_scheme}  
as well relevant literature data for similar substances are collected in Table 1. 
%\ref{Omega_table} 
In the following, we will discuss particular aspects of the optimization for the chosen target systems.

Exemplary,  in the left panel of Figure 3 
%\ref{Curves} 
details of the tuning procedure for the \ce{IrPS-Ag2} case are presented. Since  $J_{0}^{}(\omega)$ and $J_{1}^{}(\omega)$ describe systems with different numbers of electrons,  in general, their minima do not coincide. The shown example is the only case among the studied systems, where the minimum  of $J^{}_{}(\omega)$ is not well-defined due to  an almost flat region from $0.18-0.25\:\text{bohr}^{-1}_{}$, where $J_{0}^{}(\omega)$ and $J_{1}^{}(\omega)$ upon summation  compensate each other.  If the least-square function $J^{*}_{}(\omega)$  is applied instead,  a minimum at $0.19\:\text{bohr}^{-1}_{}$ can be easily located.  For all other test systems collected in Table 1, 
%\ref{Omega_table}
the minima of $J$ and $J^{*}_{}$ were very close to each other; $J^{*}_{}$  led in some cases to a decrease of  $\omega$ by $0.01-0.02\:\text{bohr}^{-1}_{}$.  Since the use of $J^{*}_{}$ was not crucial in our case, we focused on the results obtained by applying $J$, as it was the main approach of previous investigations.

\begin{table}
	\begin{threeparttable}[T]
		\caption{\label{Omega_table}
			Optimized range-separation parameters, $\omega$, 
			defined by the minumum of $J(\omega)$, eq 6,
			%\ref{Jgeneral},
			for different molecules depicted in Figure 1.
			 %\ref{First_scheme}. 
			For comparison, literature data on molecules relevant for photovoltaics 
			are provided.
		}	
		\begin{tabular*}{0.45\textwidth}{l c}
			\hline
			Compound & Optimal $\omega$ $[{\rm bohr}^{-1}]$ \\
			\hline
			IrPS & 0.18 \\
			IrPS (THF, PCM) & 0.05 \\
			IrPS + 28 THF & 0.15 \\
			\ce{[Ir(ppy)2(en)]+} & 0.18 \\
			\ce{[Ir(pox)2(bpy)]+} & 0.18 \\
			\ce{[Ir(pbox)2(bpy)]+} & 0.18\\
			\ce{[Ir(ppy)2(bpym)Ir(ppy)2]+} & 0.14 \\
			\ce{[Cu(phen$^{*}$)(xant)]+} & 0.16 \\
			\hline
			\ce{Ag2} & 0.42 \\ 
			\ce{Ag10} & 0.22 \\
			\ce{Ag20} & $\approx$0.20\tnote{a}\\
			\ce{IrPS-Ag2} & 0.18 -- 0.25 \tnote{b}\\
			\ce{IrPS-Ag10} & 0.16 \\
			\ce{IrPS-Ag20} & 0.16 \\ 
			\hline
			\ce{TEA} & 0.31 \\
			\ce{IrPS-TEA} & 0.23 -- 0.32 \tnote{c} \\
			\hline
			Fe-cat(1) & 0.24 \\
			Fe-cat(2) & 0.19 \\
			Fe-cat(3) & 0.17 \\
			IrPS-Fe-cat(1) & 0.18 \tnote{c} \\
			IrPS-Fe-cat(2) & 0.17 -- 0.18 \tnote{c}\\
			IrPS-Fe-cat(3) & 0.16 -- 0.17 \tnote{c} \\
			\hline
			\ce{C60}\cite{Refaely-prb-2011} & 0.21 \\
			phthalocyanine \cite{Refaely-prb-2011} & 0.16 \\
			Si-nanocristals (5-15 \AA) \cite{Stein-prl-2010} & 0.10 -- 0.24 \\
			coumarine dyes \cite{Stein-jcp-2009} & 0.17 -- 0.21 \\
			pentacene-\ce{C60} \cite{Minami-ijqc-2013} & 0.21 \\
			nucleobases \cite{Foster-jctc-2012} & 0.27-0.31 \\
			\hline
		\end{tabular*}
		\begin{tablenotes}
			\item [a] Because of convergence problems, only tentative a value is given;
			\item [b] using $J(\omega)$, eq 6, 
%			(Eq. \ref{Jgeneral}) 
			a flat dependence is obtained, see text and Figure 3; 
			%\ref{Curves}; 
			\item[c] different geometries are used, see text.
		\end{tablenotes}
	\end{threeparttable}
\end{table}

In the right panel of Figure 3 
%\ref{Curves}, 
selected  examples of $J(\omega)$ for \ce{IrPS} in different environments are presented. The curve for IrPS(vac) represents a case with  regular behaviour, i.e. there is a  distinct minimum.  For \ce{IrPS-Ag10} the region after $0.25\:\text{bohr}^{-1}_{}$  corresponds to a changed order of  HOMO and HOMO-1 orbitals for $N+1$-electron system. For the isolated reduced IrPS in the doublet electronic state,  the unpaired electron was previously shown to be localized on $\pi\text{(bpy)}^{*}_{}$ orbital.~\cite{Bokarev-pccp-2014}  For the \ce{IrPS-Ag10}, the internal oxidation-reduction occurs for $\omega > 0.25\: \text{bohr}^{-1}_{}$: the unpaired electron moves from $\pi\text{(bpy)}^{*}_{}$ to $\sigma^{*}_{}\text{(Ag)}$.

As IrPS forms no stable complexes, \ce{IrPS-X}, neither with TEA   nor with the iron-catalysts  \cite{Neubauer-jpcl-2014},  the relative position of constituents might be flexibly varied.  Similar to the procedure in ref\,~\cite{Neubauer-jpcl-2014} we have varied the position of X=TEA  or Fe-cat on a sphere around IrPS. Thereby we found that  choosing different locations does not have an impact on resulting $\omega$. However, changing the distance between IrPS and X  led to some variation of optimal $\omega$,  see Table 1.
%. \ref{Omega_table}.
Interestingly, the optimal $\omega$ value was found to increase  in the range   $0.23-0.32\:\text{bohr}^{-1}_{}$ with increasing the distance between IrPS and TEA from 7 to 12 \AA. While the minimal value  of $J_{0}^{}(\omega)$ remains at $0.18 \:\text{bohr}^{-1}_{}$, the minimum of $J_{1}^{}(\omega)$ occurs at larger $\omega$. The usage of $J^{*}_{}(\omega)$ instead of  $J^{}_{}(\omega)$ provides a narrower range  of $0.23-0.28\:\text{bohr}^{-1}_{}$. In passing we note that in refs\,~\cite{Livshits-pccp-2007, Karolewski-jcp-2013} it was shown that  an optimization of   $\omega$  for each geometry and subsystem can lead to size-inconsistency or unphysical potential energy curves in excited states.

For comparison, the standard \cite{Tawada-jcp-2004, Song-jcp-2007} system-independent  $\omega_{1}^{}=0.33 \, \text{bohr}^{-1}_{}$ and  $\omega_{2}^{}=0.47 \, \text{bohr}^{-1}_{}$  are  shown in both panels of Figure 3
%. \ref{Curves} 
by vertical lines. As can be seen from that figure and  Table 1,
% \ref{Omega_table},
the optimally tuned $\omega$ values for compounds  relevant in photocatalysis are substantially lower than those determined for  diatomics and small molecules.~\cite{Tawada-jcp-2004, Song-jcp-2007}  Since $\omega^{-1}$ reflects a characteristic  distance for switching between short- and  long-range parts  or, in other words, an effective electron screening  (delocalization) length, previously optimal $\omega$ values were  found to decrease with increasing system size and  conjugation length.~\cite{Stein-jcp-2009,  Korzdorfer-jcp-2011, Stein-prl-2010,  Karolewski-jcp-2011, Refaely-prb-2011,  Sears-jcp-2011, Salzner-jctc-2011} However, in some cases the dependence was not monotonous  and strongly varied for systems with different electronic  structure.~\cite{Refaely-prb-2011} In the present study, the Ir(III) complexes have various sizes, being  smallest for \ce{[Ir(ppy)2(en)]+} and largest for \ce{[Ir(ppy)2(bpym)Ir(ppy)2]+}. However, only a minor  dependence of $\omega$ on the size of ligands around the central Ir atom and the size of molecules (silver clusters, iron carbonyls) bound to the PS was found (cf. Table 1).
%. \ref{Omega_table}).

\subsection{Effect of Solvation}
\begin{figure*}[t]
	\includegraphics[width=\textwidth]{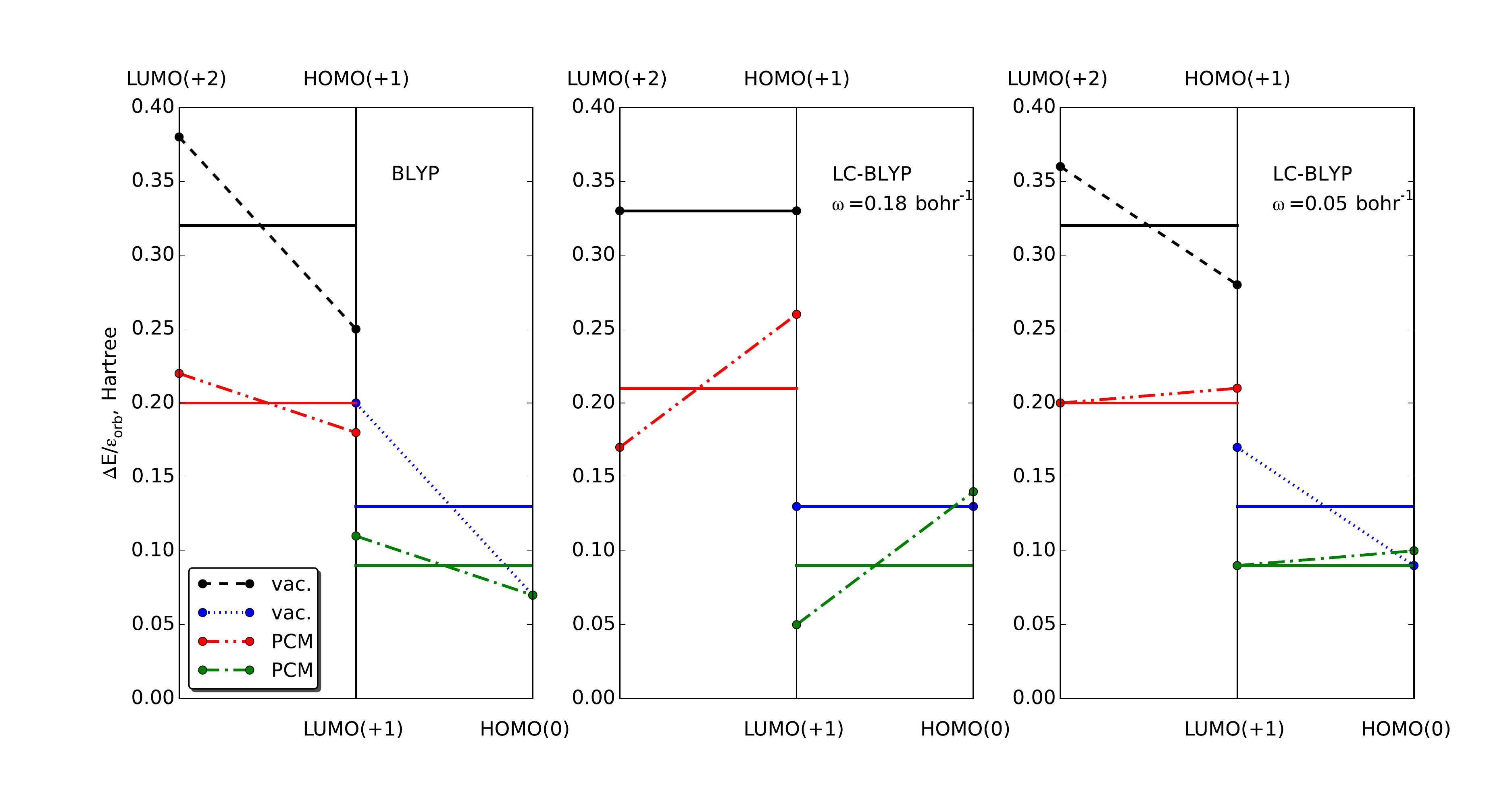}
	\caption{
		\label{Janak}
		$\Delta \text{SCF}$ redox energies $\Delta \text{E}$ (dashed lines) and orbital eigenvalues $-\varepsilon_{\text{HOMO/LUMO}}^{}$ (solid lines) for different charge states (+2,+1, and 0) of IrPS computed with pure BLYP and LC-BLYP with  $\omega$ optimization in vacuum  and PCM solvent. The eigenvalue of the $\beta$ spin-orbital was used for LUMO(+2) all other values correspond to the $\alpha$ spin-orbital.
	}
\end{figure*}
By requiring Koopmans' theorem to be satisfied simultaneously for systems with $N$ and $N+1$ electrons where HOMO and LUMO are on different fragments, we automatically improve the description of electron transfer. This implies that the long-range corrected functional with optimized range-separation parameter should have a  predictive power to study also redox properties. To describe redox (electron transfer) reactions in photocatalysis, solvents effects need to be included. However, to the best of our knowledge, so far no $\omega$-tuning in the presence of a solvent was reported. 
The green curve with square symbols in Figure 3 
%\ref{Curves} 
shows  $J(\omega)$ for \ce{IrPS}, solvated via the PCM model (THF solvent). This yields a very small optimal $\omega=0.05\:\text{bohr}^{-1}_{}$. Similar computations have been done for TEA and Fe-cat(3), resulting in values 0.10 and 0.05 $\text{bohr}^{-1}_{}$, respectively. Note that such a  small value of $\omega$ implies that there is almost no long-range correction from exact  HF exchange. Interestingly, the inclusion of 28 explicit THF molecules around IrPS, which corresponds to one solvation shell, led to $\omega=0.15\:\text{bohr}^{-1}_{}$ similar to vacuum case. 

According to Janak's theorem \cite{Janak-prb-1978} the redox energy upon changing the number of electrons from $N-1$ to $N$ can be obtained in integral form as 
\begin{equation}
\Delta E = \int\limits_{N-1}^{N}{\varepsilon(n)\, dn}.
\end{equation}
In the absence of self-interaction and delocalization errors, the orbital energies $\varepsilon$ stay constant between integer numbers of electrons, thus satisfying the condition of $\omega$ optimization, eq 2. 
%\ref{eq:IP}. 
In Figure 4 redox and HOMO/LUMO energies are shown for different functionals and optimization conditions. For the tuned $\omega$, $\Delta E=-\varepsilon_{\text{HOMO}}^{}(N)=-\varepsilon_{\text{LUMO}}^{}(N-1)$ are given  in central panel of Figure 4 
%\ref{Janak} 
for IrPS in vacuum. When using the conventional BLYP functional this condition is not satisfied (left panel). However, if we include implicit PCM solvation the jump in energies of HOMO($N$) and LUMO($N-1$) is mitigated by the response of the polarizable continuum. This leads to much better estimates of redox energies obtained with conventional functionals with PCM as compared with the vacuum.~\cite{Chen-jpca-2013} In other words, in terms of the derivative discontinuity the PCM solvent effectively leads to a partial mitigation of the $E_{\rm XC}^{}$ deficiencies. 

That is why one needs to include much less exact exchange into the LC-BLYP functional (smaller $\omega$) to satisfy Koopmans' theorem and  derivative discontinuity condition as is seen from Figure 3 
%\ref{Curves} 
and Table 1. 
%\ref{Omega_table}. 
The orbital eigenvalues in the right panel of Figure 4 
%\ref{Janak} 
for the solvent case are almost constant, the deviations being probably due to a residual delocalization error. Importantly, the $\omega$ value optimized for the vacuum does not fulfil the condition of constant orbital energies in solvent and vice versa. Further examples of Fe-cat(3) and TEA leading to the same conclusion can be found in the Supporting Information.

\subsection{Triplet Stability}

\begin{figure}[t] 
	\includegraphics[width=0.55\textwidth]{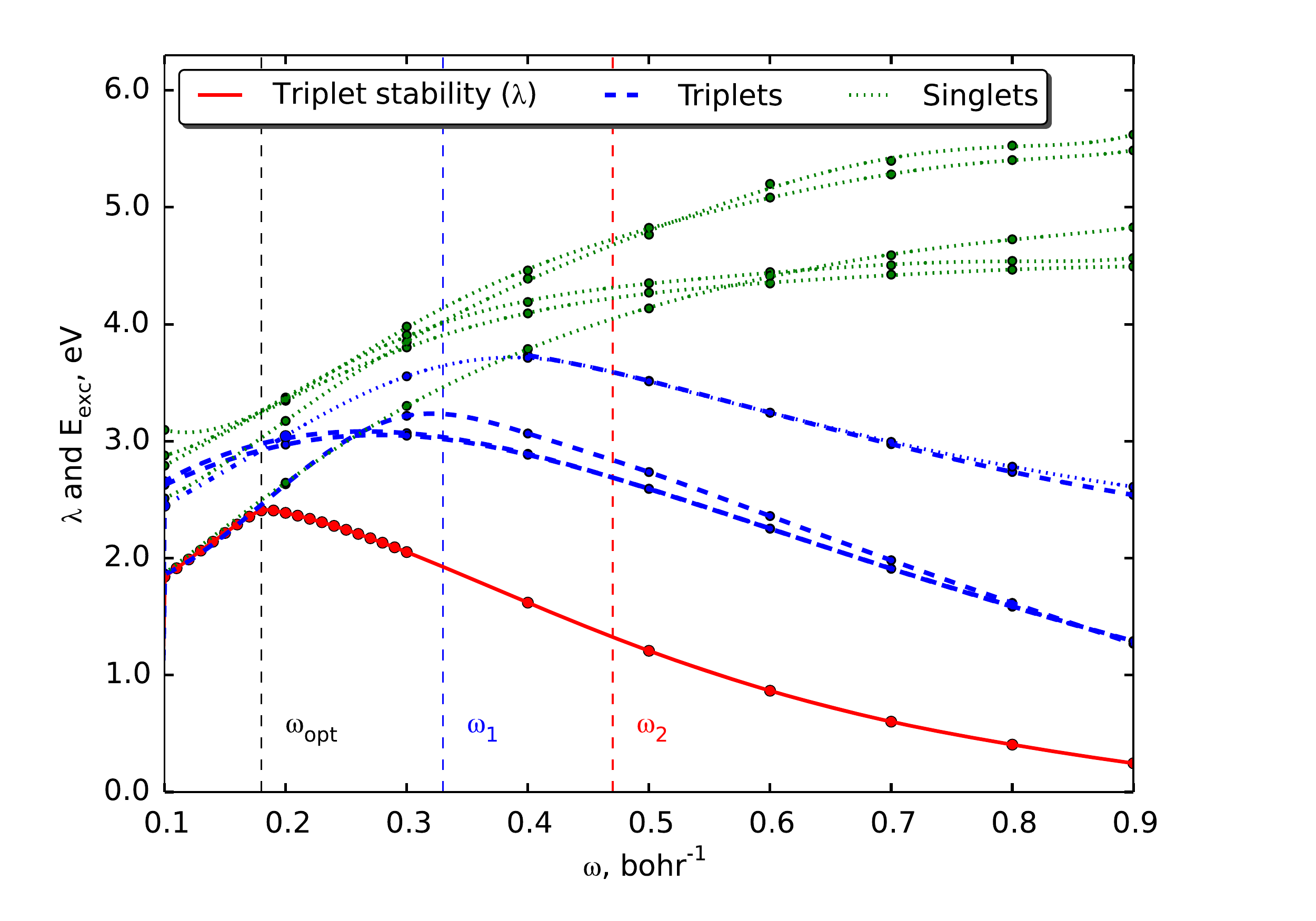}
	\caption{\label{Stab}
		Energies of the lowest five singlet and triplet excited states 
		and the lowest eigenvalue of the ground state stability matrix 
		($\lambda$) in dependence on the range 
		separation parameter $\omega$ for IrPS in vacuum.}
\end{figure}
The nature and energetic position of the lowest triplet state of PSs need to be calculated very accurately since this state plays a key role in photoprocesses as a dominant long-living and emitting excited state. However, the calculation of triplet  states represents a challenge for the TDDFT approach with long-range corrected hybrid functionals because of  symmetry breaking instabilities of the ground  state solution.~\cite{Bauernschmitt-cpl-1996, Cizek-jcp-1967} Typically, the larger amount of exact exchange leads to more pronounced problems with instabilities of the ground state wave function. This results in a divergence of TDDFT excitation energies and an imaginary energy of  the lowest triplet state.~\cite{Casida-jcp-2000} As shown in ref\,~\cite{Sears-jcp-2011}, the non-empirical tuning of $\omega$ could avoid the instabilities  through the stabilizing the ground state solution, but this issue should be carefully analyzed for the particular system under study. Since the instability problems are generally more pronounced for  the HF method than for DFT,~\cite{Sears-jcp-2011, Bauernschmitt-cpl-1996}  the application of a smaller value for $\omega$  should result in a more stable ground-state solution. In Figure 5, 
%\ref{Stab} 
the dependency of the smallest eigenvalue of triplet stability matrix \cite{Bauernschmitt-cpl-1996} on the range-separation parameter, $\lambda(\omega)$,  for the case of IrPS is given; the energies of lowest triplet and singlet energies are provided as well.

It can be clearly seen that for the present system the maximum of stability $\lambda^{}_{\rm max}$ corresponds to the optimal $\omega$ value. The $\lambda$ value is notably higher  for the standard values $\omega_{1,2}$.~ \cite{Tawada-jcp-2004, Song-jcp-2007} This implies that LC-BLYP with properly tuned range-separation parameter ensures a more reliable estimate of triplet state energies. The singlet state energies systematically rise with the increase of the exact exchange part, which is a common behaviour for TDDFT.~\cite{Bokarev-jcp-2012} For triplet states, this rising is compensated by the decrease of energies  upon increase of instability resulting in  maxima around  $0.25-0.40 \, \text{bohr}^{-1}_{}$.

\subsection{Electronic Absorption Spectra}

\begin{figure*}[t]
	\includegraphics[width=0.95\textwidth]{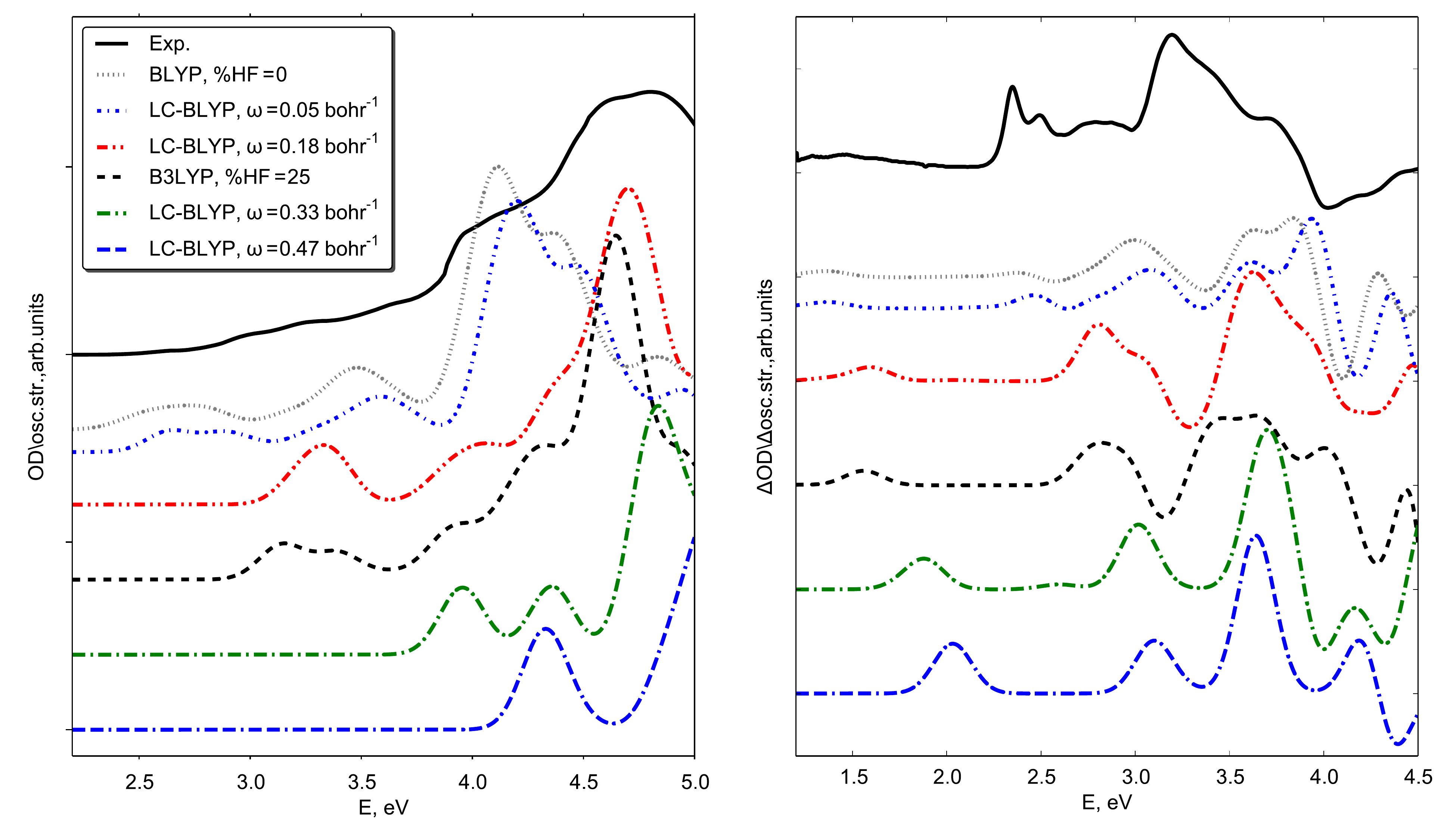}
	\caption{
		\label{Spectra}
		Left panel: Experimental (optical density, OD) and theoretical (TDDFT, oscillator strength) electronic excitation spectra.  Right panel:  Difference absorption spectra (reduced species - oxidized species) obtained theoretically and by transient absorption experiments		 of IrPS in THF-solution. Experimental spectra were measured by Lochbrunner et al.  and had been published together with the calculated  BLYP and B3LYP spectra in refs\,~\cite{Bokarev-jcp-2012, Bokarev-pccp-2014}. All calculations included THF within the PCM.
	}
\end{figure*}

A comparative analysis of excitation and differential absorption spectra of IrPS in THF calculated by means of local (BLYP), hybrid (B3LYP), and long-range corrected LC-BLYP with standard and both optimal (for vacuum and THF) $\omega$ values is provided in Figure 6. 
%\ref{Spectra}. 
In the TDDFT calculations, the 80 lowest singlet-singlet states have been included and assigned. A phenomenological broadening  by a Gaussian lineshape with a width of 0.1 eV was assumed. 

Overall, the features of the absorption spectrum are well reproduced by B3LYP and LC-BLYP with optimal  $\omega = 0.18 \: \text{bohr}^{-1}_{}$. However, absorption and difference spectra could only be properly assigned simultaneously using LC-BLYP ($\omega = 0.18 \: \text{bohr}^{-1}_{}$).~\cite{Bokarev-pccp-2014} It should be noted, that besides the overall shape of spectra, which can be described by the conventional hybrid functional, the quality of description of CT states and band gap characteristics is only assured for the  long-range functional with the properly tuned $\omega$ parameter. The excessive amount of HF exchange (as in LC-BLYP with both standard $\omega$ values) as well as complete absence of exact exchange (as in BLYP) led to shapes of absorption and difference spectra of IrPS deviating from experiments. Naturally, the results of LC-BLYP with optimal $\omega$ tuned for PCM are very close to those of pure BLYP because of only small amount of exact exchange is included. Hence, LC-BLYP with $\omega = 0.05 \: \text{bohr}^{-1}_{}$ optimized in a PCM solvent model  does not agree with experiment, similar to the parent BLYP functional which is known for wrong prediction of energies of CT states.

The latter observation deserves some more comments. In view of the derivative discontinuity, $\omega = 0.05 \: \text{bohr}^{-1}_{}$  obtained for the PCM model gives an improved description as compared to the gas phase value (see Figure 4, 
%\ref{Janak}, 
right panel). In contrast, using the gas phase $\omega = 0.18 \: \text{bohr}^{-1}_{}$ together with a PCM model yields an ``overscreening'' (see Figure 4,
%\ref{Janak}, 
middle panel). At this point, one should  note  that under the influence of a polar solvent the energies of CT and local states shift differently from their vacuum values because of substantially different dipole moments. That is why, to correctly predict the electronic absorption spectrum in solution, one needs to use  $\omega$ tuned in vacuum and to include  solvent effects via PCM. In other words, we observe that, when using the PCM model, for obtaining a reasonable agreement between measured and calculated absorption spectra one has to sacrifice the derivative discontinuity condition.

%-------------------------------------------------------------
\section{Conclusions}
\label{sec:Conc}
%-------------------------------------------------------------
%
The present investigation of molecules and complexes relevant to photochemical water splitting points to the need for a system-specific optimization of the range-separation parameter within the LC-BLYP approach. It   provides an improved band gap characteristics, increased ground state stability, and a correct asymptotic behaviour of CT states. The actual optimized value, $\omega = 0.18 \: \text{bohr}^{-1}_{}$, was found to be nearly the same for various Ir and Cu-based PS as well as for IrPS complexes with silver clusters and iron carbonyls. Using the $\omega$ optimization it was shown that  the tuning for supermolecular compound systems, where an electron relay between constituents is of interest, gives an improvement of the energies of HOMO and LUMO orbitals localized on different fragments. Thus, this  approach is in principle suitable for accurate modelling of redox molecular properties. 

The optimization of the range-separation parameter is usually performed for molecules in vacuum. Here, for the first time, an optimization in the presence of a solvent described by the PCM model has been done. For the used $\Delta$SCF approach this yielded some unexpected results, namely that the optimal $\omega$ tends to zero. As a consequence there is almost no long-range correction and the quality, e.g. of electronic excitation spectra, corresponds to that of the BLYP functional.  Since CT and long-range properties are of primary interest for photocatalysis, as a conclusion it is recommended to use  $\omega$ optimized in vacuum together with PCM. This seems to  provide a better description of CT  processes and electronic absorption spectra, even though the  formal conditions imposed by the derivative discontinuity are violated. Therefore, if redox reactions are studied, the combination $\Delta$SCF/PCM is not recommended. It either leads  to derivative discontinuity problems or to a wrong asymptotic behaviour of the  optimized functional. Needless to say, that this procedure should not be applied without careful diagnostics, e.g., as provided by Figure 4.
%\ref{Janak}

\begin{acknowledgement}
	This work has been  supported by the BMBF within the project ``Light2Hydrogen'' (Spitzenforschung und Innovation in den Neuen L\"andern), by the European Union (European Social Funds, ESF) within the project ``PS4H'' and by the Ministry for Education, Science and Culture of Mecklenburg-Vorpommern. 
	\end{acknowledgement}

\begin{suppinfo}
	Full data on $\Delta$SCF redox energies and orbital eigenvalues for IrPS, TEA, and Fe-cat(3).
\end{suppinfo}

%\bibliography{omega_short}

\providecommand{\latin}[1]{#1}
\providecommand*\mcitethebibliography{\thebibliography}
\csname @ifundefined\endcsname{endmcitethebibliography}
{\let\endmcitethebibliography\endthebibliography}{}

%\end{document}	
\clearpage
%\noindent
\begin{minipage}{\textwidth}

%----------------------------------------------------------
\section*{Supporting Information}
\label{sec:SI}
%----------------------------------------------------------
%
%\begin{figure}
	\label{Janak_full}
	\includegraphics[width=\textwidth]{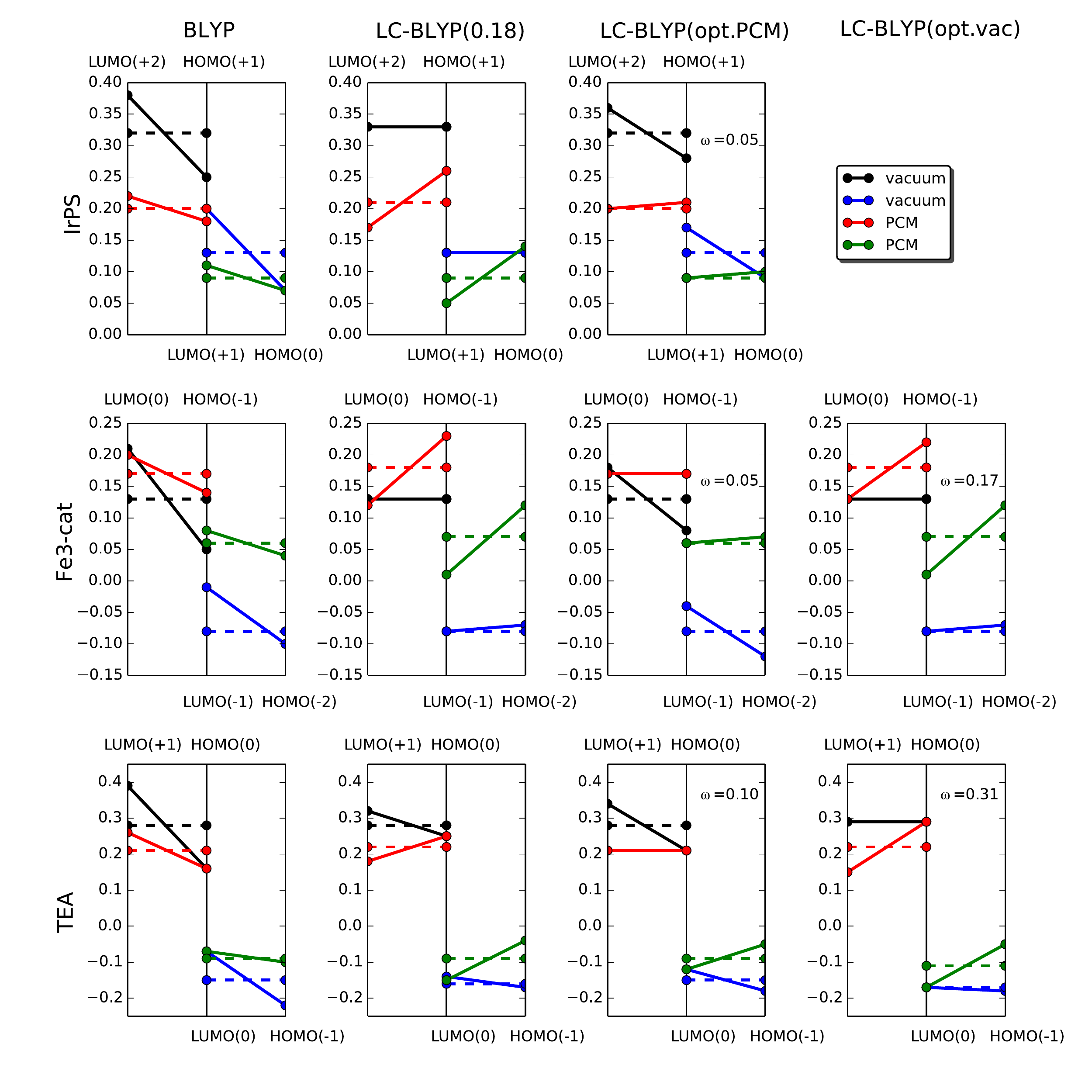}
Fig. S1. $\Delta \text{SCF}$ redox energies ($\Delta \text{E}$) and orbital eigenvalues ($-\varepsilon_{\text{HOMO/LUMO}}^{}$) for different charged states of IrPS, Fe-cat(3), and TEA computed with pure BLYP and LC-BLYP with different $\omega$ in vacuum and PCM solvent (THF). Values of $\omega$ are optimized for IrPS in vacuum ($0.18 \: \text{bohr}^{-1}_{}$) as well as for each compound in vacuum and THF (given at the corresponding panels)
%\end{figure}
\end{minipage}
%

%\twocolumn[]
\end{document}